\documentclass[aps,prd,twocolumn,showpacs,preprintnumbers,amsmath,amssymb,groupedaddress]{revtex4}
\usepackage{tipa}
\usepackage{graphicx}
\usepackage{txfonts}

\begin{document}


\title{Polarizable vacuum analysis of the gravitational metric tensor}


\author{Xing-Hao Ye}
\email[Electronic address: ]{yxhow@163.com}

\affiliation{Department of Applied Physics, Hangzhou Dianzi
University, Hangzhou 310018, China}


\date{\today}

\begin{abstract}
The gravitational metric tensor implies a variable dielectric tensor
of vacuum around gravitational matter. The curved spacetime in
general relativity is then associated with a polarizable vacuum. It
is found that the number density of the virtual dipoles in vacuum
decreases with the distance from the gravitational centre. This
result offers a polarizable vacuum interpretation of the
gravitational force. Also, the anisotropy of vacuum polarization is
briefly discussed, which appeals for observational proof of
anisotropic light propagation in a vacuum altered by gravitational
or electromagnetic field.
\end{abstract}

\pacs{31.30.jf, 04.20.Cv, 42.25.Bs}

\maketitle


\section{Introduction}
\label{}

Ever since the establishment of general relativity, the search for
the true nature of gravitational field has never ceased. In
Einstein's geometric description, the gravitational force is
regarded as an effect of curved spacetime. Nevertheless, Einstein
himself proposed in 1920 that ``according to the general theory of
relativity space is endowed with physical qualities''
\cite{Einstein1920}. Meanwhile, Eddington suggested that the
deflection of light in a curved spacetime can be imaged as a
refraction effect in a flat spacetime \cite{Eddington1920}. And
Wilson related this refractive index to the variable specific
inductive capacity of the ``ether'' near matter, i.e., suggested an
electromagnetic theory of gravitation \cite{Wilson1921}. Later, in
1957, Dicke \cite{Dicke1957}, pointed out that vacuum polarization
\cite{Milonni1994,Peskin2006} can account for such ``ether''.
Recently, Puthoff reemphasized such an interpretation of
gravitational field, and suggested an exponential form of the vacuum
dielectric constant or vacuum refractive index \cite{Puthoff2002}.
Vlokh further investigated the constitutive coefficients that could
characterize the space as matter, and proposed the concept of
optical-frequency dielectric impermeability perturbed by a
gravitational field \cite{Vlokh2004,Vlokh2005,Vlokh2007}. The
dielectric-like vacuum interpretation of gravitational field has led
to successful analysis of the gravitational effects such as the
gravitational red shift, the light deflection, the time delay, the
gravitational lensing, and so on
\cite{Dicke1957,Puthoff2002,Vlokh2004,Vlokh2005,Vlokh2007,Puthoff2005,Vlokh2006,Ye2008-01,Ye2008-02}.

In this paper, the relation between the gravitational field and the
quantum vacuum will be discussed further. First, the gravitational
metric tensor will be analysed and a corresponding graded refractive
index in the flat spacetime will be deduced. This refractive index
implies a variable dielectric tensor of the gravitational space,
which is essentially related to the polarization of vacuum. Then,
through the analysis of the intensity of vacuum polarization, the
number density of the virtual charge pairs in vacuum will be found
to be a function of the gravitational mass $M$ and the distance $r$.
This analysis leads to a polarizable vacuum interpretation of the
gravitational force. Finally, the anisotropic propagation of light
in a vacuum will be discussed.

\section{From metric tensor to dielectric tensor}
\label{}

The general relativity states that spacetime will be curved by the
gravitational matter. A curved spacetime can be described by the
four-dimensional Riemann space, where the velocity of light in
vacuum is defined as a constant $c$. The interval $ds=cd\tau$
($\tau$ is the proper time) between the two adjacent points in this
space is
\begin{equation}
ds^2=-g_{\mu\nu} dx^\mu dx^\nu,
\end{equation}
where $x^\mu$ is the coordinates of the point, for example,
$x^0=ict$ ($t$ denotes the coordinate time), $x^1=x$, $x^2=y$,
$x^3=z$, and $g_{\mu\nu}$ is the metric tensor
\begin{equation}
g_{\mu\nu}=
\left(
  \begin{array}{cccc}
    g_{00} &  g_{01} &  g_{02} &  g_{03} \\
    g_{10} &  g_{11} &  g_{12} &  g_{13}  \\
    g_{20} &  g_{21} &  g_{22} &  g_{23}  \\
    g_{30} &  g_{31} &  g_{32} &  g_{33}  \\
  \end{array}
\right).
\end{equation}

For a static and spherically symmetric gravitational matter with
mass $M$, the Schwarzschild exterior solution gives the metric
\cite{Weinberg1972}
\begin{equation}
ds^2=-B(R)(icdt)^2-A(R)dR^2-R^2(d\theta^2+\sin^2\theta d\varphi^2),
\end{equation}
where $B(R)=(1-2GM/Rc^2)$, $A(R)=(1-2GM/Rc^2)^{-1}$, and $R$,
differing from the radius $r$ in flat spacetime, is the radial
coordinate of the metric, $G$ is the gravitational constant,
$\theta$ and $\varphi$ have the same meaning as those in the usual
spherical coordinates. The above equation gives the metric tensor
\begin{equation}
g_{\mu\nu}= \left(
  \begin{array}{cccc}
    B(R) & 0 & 0 & 0 \\
    0 & A(R) & 0 & 0 \\
    0 & 0 & R^2 & 0 \\
    0 & 0 & 0 & R^2\sin^2\theta \\
  \end{array}
\right).
\end{equation}

In a rectangular coordinate system of flat spacetime, if we set
$dx^0=ic'dt$ (where $c'$, differing to the constant velocity $c$, is
the velocity of light in vacuum at distance $r$ from the
gravitational centre in flat spacetime), $dx=dr$, $dy=rd\theta$,
$dz=r\sin\theta d\varphi$, then Eq.\ (3) can be rewritten as
\begin{equation}
ds^2=-B(R)\left(\frac{c}{c'}\right)^2(ic'dt)^2-A(R)\left(\frac{dR}{dr}\right)^2
dx^2-\left(\frac{R}{r}\right)^2(dy^2+dz^2),
\end{equation}
and the metric tensor will be
\begin{equation}
g'_{\mu\nu}= \left(
  \begin{array}{cccc}
    B(R)\left(\frac{c}{c'}\right)^2 & 0 & 0 & 0 \\
    0 & A(R)\left(\frac{dR}{dr}\right)^2 & 0 & 0 \\
    0 & 0 & \left(\frac{R}{r}\right)^2 & 0 \\
    0 & 0 & 0 & \left(\frac{R}{r}\right)^2 \\
  \end{array}
\right).
\end{equation}

Since the metric tensor of a flat spacetime satisfies
$g'_{00}=g'_{11}=g'_{22}=g'_{33}$, we have
\begin{equation}
\sqrt{B(R)}\frac{c}{c'}=\frac{R}{r},
\end{equation}
\begin{equation}
\sqrt{A(R)}\frac{dR}{dr}=\frac{R}{r}.
\end{equation}

Considering $R\rightarrow r$ at infinity, Eq.\ (8) gives
\begin{equation}
r=\frac{R}{2}\left(\sqrt{1-\frac{2GM}{Rc^2}}+1-\frac{GM}{Rc^2}\right),
\end{equation}
or
\begin{equation}
R=r\left(1+\frac{GM}{2rc^2}\right)^2.
\end{equation}

Combining Eqs.\ (7) and (9), we get the refractive index
\begin{equation}
n=\frac{c}{c'}=\frac{1}{\frac{1}{2}
\left(\sqrt{1-\frac{2GM}{Rc^2}}+1-\frac{GM}{Rc^2}\right)\sqrt{1-\frac{2GM}{Rc^2}}},
\end{equation}
or through Eqs.\ (7) and (10), we get
\begin{equation}
n=\left(1+\frac{GM}{2rc^2}\right)^3
\left(1-\frac{GM}{2rc^2}\right)^{-1}.
\end{equation}

The above results are in agreement with those obtained through an
optical way \cite{Ye2008-02}.

If we set a new tensor $\eta'_{\mu\nu}$, each of its elements is the
square root of the corresponding element of $g'_{\mu\nu}$, then we
have
\begin{equation}
\frac{\eta'_{\mu\nu}}{\sqrt{1-\frac{2GM}{Rc^2}}}= \left(
  \begin{array}{cccc}
    \frac{c}{c'} & 0 & 0 & 0 \\
    0 & n & 0 & 0 \\
    0 & 0 & n & 0 \\
    0 & 0 & 0 & n \\
  \end{array}
\right).
\end{equation}

In weak field approximation, $n=\exp\ (2GM/rc^2)$, the tensor
becomes
\begin{equation}
\frac{\eta'_{\mu\nu}}{\sqrt{1-\frac{2GM}{Rc^2}}}= \left(
  \begin{array}{cccc}
    \frac{c}{c'} & 0 & 0 & 0 \\
    0 & \exp{\left(\frac{2GM}{rc^2}\right)} & 0 & 0 \\
    0 & 0 & \exp{\left(\frac{2GM}{rc^2}\right)} & 0 \\
    0 & 0 & 0 & \exp{\left(\frac{2GM}{rc^2}\right)} \\
  \end{array}
\right).
\end{equation}

The above result implies a variable dielectric tensor
$\varepsilon'_0$ of vacuum around gravitational matter:
\begin{equation}
\varepsilon'_0= \varepsilon_0\left(
  \begin{array}{ccc}
     \exp{\left(\frac{2GM}{rc^2}\right)} & 0 & 0 \\
    0 & \exp{\left(\frac{2GM}{rc^2}\right)} & 0 \\
    0 & 0 & \exp{\left(\frac{2GM}{rc^2}\right)} \\
  \end{array}
\right),
\end{equation}
where $\varepsilon_0$ is the constant permittivity of vacuum without
the influence of gravitational matter.

Eqs.\ (14) and (15) correspond to a graded refractive index of the
vacuum
\begin{equation}
n=\exp{\left(\frac{2GM}{rc^2}\right)},
\end{equation}
and a variable light velocity in gravitational field
\begin{equation}
c'=c\exp{\left(-\frac{2GM}{rc^2}\right)}.
\end{equation}

\section{Polarizable vacuum analysis of the tensor}
\label{}

It is noticed that the electric displacement vector \textbf{D} in a
dielectric medium can be written as
\begin{equation}
\mathbf{D}=\mathbf{P}+\varepsilon_0
\mathbf{E}=\mathbf{P}+\mathbf{P'},
\end{equation}
where \textbf{P} represents the polarization of dielectric medium,
$\mathbf{P'}=\varepsilon_0 \mathbf{E}$ is the polarization of vacuum
\cite{Dicke1957,Puthoff2002,Ye2009-01}, \textbf{E} denotes the
electric field intensity.

For the vacuum around gravitational matter, the variable dielectric
tensor expressed by Eq.\ (15) indicates that the intensity of vacuum
polarization will be
\begin{equation}
\mathbf{P'}=\varepsilon_0 ' \mathbf{E}=\varepsilon_0
\left(\frac{2GM}{rc^2}\right) \mathbf{E}.
\end{equation}

As we know that, the intensity of dielectric polarization is
\begin{equation}
\mathbf{P}=\frac{n_{\textnormal{p}} e^2}{k} \mathbf{E},
\end{equation}
where $n_{\textnormal{p}}$ is the number density of electric dipoles
in the dielectric medium, $e$ is the electronic charge, $k$ is the
stiffness factor. Similarly, we suppose the intensity of vacuum
polarization to be
\begin{equation}
\mathbf{P'}=\frac{n_{\textnormal{p}'} q'^2}{k} \mathbf{E},
\end{equation}
where $n_{\textnormal{p}'}$ is the number density of the virtual
charge pairs $\pm q'$ in the vacuum at a distance $r$ from the
gravitational mass $M$. Thus we have
\begin{equation}
\frac{n_{\textnormal{p}'} q'^2}{k}=\varepsilon_0
\left(\frac{2GM}{rc^2}\right),
\end{equation}
or
\begin{equation}
n_{\textnormal{p}'}=\frac{\varepsilon_0}{q'^2}k
\left(\frac{2GM}{rc^2}\right).
\end{equation}

In general relativity, a clock at a distance $r$ from the
gravitational centre is slower than that at infinity. This effect of
time dilatation can be understood as the frequency reduction of the
virtual dipoles in gravitational field, that is
\begin{equation}
\omega=\sqrt{\frac{k}{m'}}=\omega_0\exp{\left(-\frac{GM}{rc^2}\right)},
\end{equation}
where $m'$ is the equivalent mass of a virtual dipole at distance
$r$, $\omega_0$ is the corresponding angular frequency at infinity.
Considering the energy ratio of a virtual dipole at distance $r$ to
that at infinity
\begin{equation}
\frac{E}{E_0}=\frac{\hbar\omega}{\hbar\omega_0}=\exp
\left(-\frac{GM}{rc^2}\right)=\frac{m'c'^2}{m'_0c^2},
\end{equation}
we have
 \begin{equation}
m'=m'_0 \exp \left(\frac{3GM}{rc^2}\right),
\end{equation}
where $m'_0$ is the corresponding equivalent mass at infinity.

Eqs.\ (24) and (26) give the relation
\begin{equation}
k=m'_0\omega_0^2 \exp \left(\frac{GM}{rc^2}\right)=k_0 \exp
\left(\frac{GM}{rc^2}\right),
\end{equation}
where $k_0$ is the stiffness factor at infinity. Therefore
\begin{equation}
n_{\textnormal{p}'}=n_{\textnormal{p}'0} \exp
\left(\frac{3GM}{rc^2}\right),
\end{equation}
where $n_{\textnormal{p}'0}=\varepsilon_0 k_0/q'^2$ is the number
density of the virtual dipoles at infinity. This result is very
similar to the number density distribution of gas molecules in a
gravitational field:
\begin{equation}
n_{\textnormal{gas}}=n_{\textnormal{gas}0} \exp
\left(-\frac{E_\textnormal{p}}{k_\textnormal{B}T}\right)=n_{\textnormal{gas}0}
\exp \left(\frac{3GM}{r\overline{v^2}}\right),
\end{equation}
where we have used the relations $E_\textnormal{p}=-\ GMm/r$ (the
potential energy of a gas molecule) and
$m\overline{v^2}/2=3k_\textnormal{B}T/2$ (the average translational
kinetic energy of a molecule).

Eq.\ (28) can be rewritten as
\begin{equation}
\frac{n_{\textnormal{p}'}}{n_{\textnormal{p}'0}}=
\frac{N/V}{N/V_0}=\frac{l_0^3}{l^3}=\exp
\left(\frac{3GM}{rc^2}\right),
\end{equation}
where $N$ is the number of virtual dipoles in a volume of
$V_0=l_0^3$ at infinity or in a volume of $V=l^3$ at a distance $r$
from the gravitational centre. So we have the length relation
\begin{equation}
l=l_0\exp{\left(-\frac{GM}{rc^2}\right)},
\end{equation}
which increases our understanding of the length contraction effect
in gravitational field. It is the effects of time dilatation
described by Eq.\ (24) and length contraction described by Eq.\ (31)
that leads to the slowing down of the light velocity $c'$ in a
gravitational field as expressed in Eq.\ (17), which, in general
relativity, is regarded as a constant light velocity $c$ in a curved
spacetime.

\section{Gravitational force}
\label{}

Eq.\ (28) tells that the number density of the virtual dipoles in
vacuum decreases with the distance $r$ from the gravitational
centre. This result indicates that there can be a polarizable vacuum
interpretation of the gravitational force.

\begin{figure}
\includegraphics[width=2.7in]{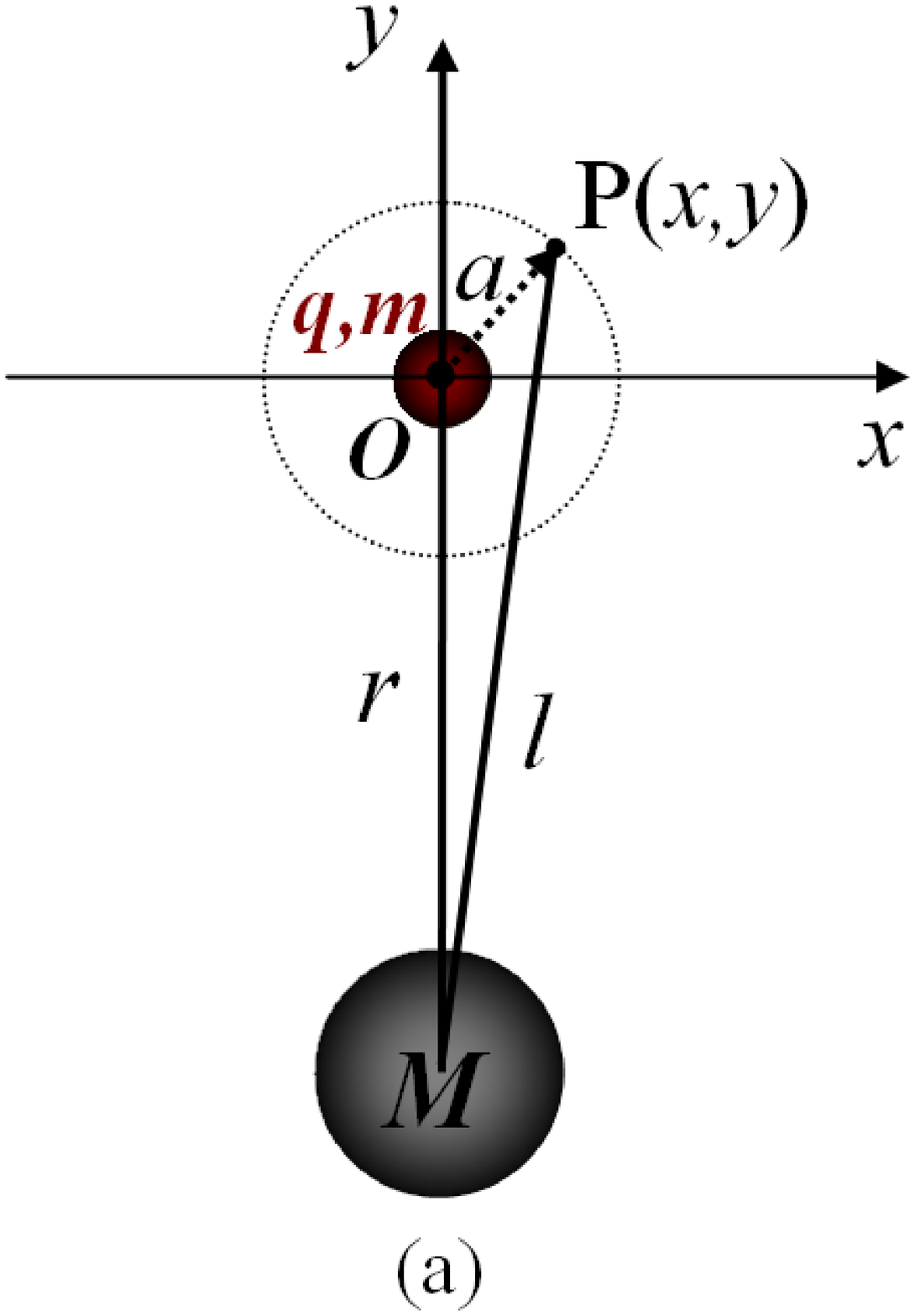}
\includegraphics[width=2.7in]{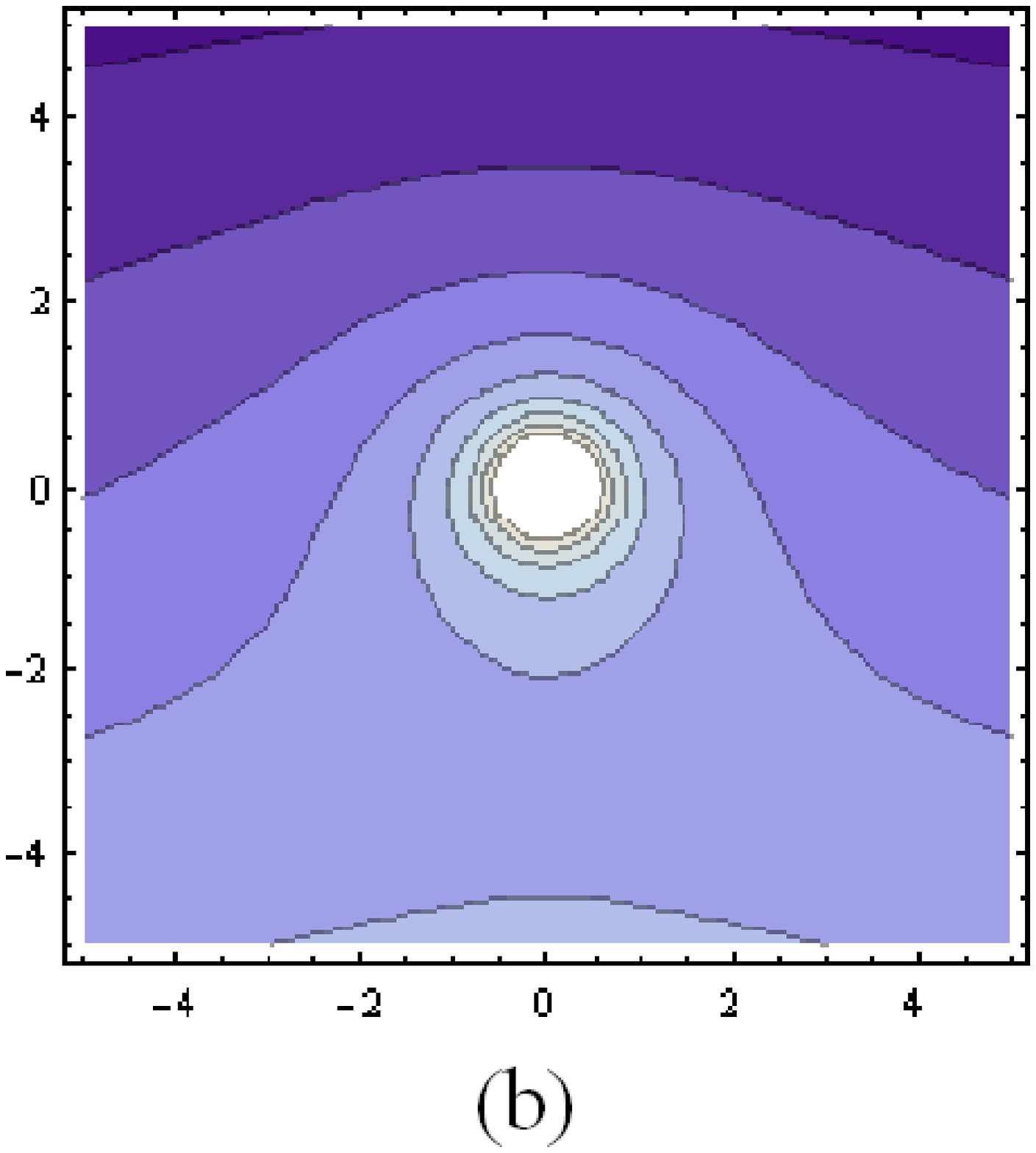}
\centering \caption{\label{fig01-b} Consider the number density of
the virtual dipoles around a $q$-charged particle in a gravitational
field.}
\end{figure}

First we consider a $q$-charged particle located at a distance $r$
from a gravitational mass $M$ as shown in Fig.\ 1 (a). According to
Eq.\ (28), the number density of the virtual dipoles around the
charged particle is
\begin{eqnarray}
n_{\textnormal{p}'}=n_{\textnormal{p}'0} \exp
\left[\frac{3G}{c^2}\left(\frac{M}{l}+\frac{m}{a}\right)\right]\ \ \ \ \ \ \ \ \ \ \ \ \ \ \ \ \ \ \ \ \ \ \ \ \ \ \ \  \ \ \ \ \ \nonumber\\
=n_{\textnormal{p}'0}
\exp\left[\frac{3G}{c^2}\left(\frac{M}{\sqrt{x^2+(y+r)^2}}+\frac{m}{\sqrt{x^2+y^2}}\right)\right],
\end{eqnarray}
where $m$ is the mass of the particle, $a$ is the distance from the
particle to a nearby point P($x$,$y$), $l$ is the distance from the
point to the gravitational matter $M$. The corresponding contour map
of $n_{\textnormal{p}'}$ is shown in Fig.\ 1 (b), where a relatively
central contour line has higher value of number density than an
outer one. From the figure we know that, for a given distance $a$,
the virtual dipoles in vacuum below the charged particle ($y<0$) are
denser than that above the charged particle ($y>0$). Then there is a
slightly higher probability for the charged particle to be coupled
or ``attracted'' by the virtual dipoles at the lower side, which
leads to the downward movement of the charged particle, or in other
words, leads to the gravitational force on the particle.

For an electrically neutral particle or neutral matter, its inner
distribution of positive and negative charges will in the same way
lead to the effect of gravitational force.

\section{Anisotropic propagation of light in vacuum}
\label{}

Once the gravitational metric tensor is associated with the
dielectric tensor of the vacuum, it is natural to have the idea that
the tensor may be in a more general form under certain conditions,
that is
\begin{equation}
\varepsilon'_0= \varepsilon_0\left(
  \begin{array}{ccc}
     a & 0 & 0 \\
    0 & b & 0 \\
    0 & 0 & c \\
  \end{array}
\right).
\end{equation}
If so, the vacuum in a specific gravitational field will exhibit its
anisotropic characteristics in polarization. For example, a vacuum
in a nonstatic gravitational field may no longer keep the relation
of $a$=$b$=$c$ and may behave somewhat like a birefringent crystal.
In such a case, there may be an anisotropic propagation of light in
the gravitational field.

The gravitational birefringence in a vacuum is an analogy of the
photoelastic effect. In the latter case, an isotropic material turns
to be anisotropic and shows its birefringent characteristic in light
propagation when a stress is applied to the material.

Similarly, if altered by a specific electric or magnetic field, an
isotropic vacuum may be turned into an anisotropic one just as an
isotropic dielectric medium will be. In such a case, there may be
birefringence observed in the altered vacuum as the electro-optic or
magneto-optic effect observed in a dielectric medium.

It is delightful that the anisotropic propagation of light in vacuum
has attracted both theoretical and experimental interests in recent
years \cite{Preuss2005,Zavattini2006}. It is hoped that substantial
observational proof will be found for the birefringence of light in
a gravity-or-electromagnetism-influenced vacuum.

\section{Conclusions}
\label{}

The curved spacetime in general relativity is described by a
gravitational metric tensor. It is found that this tensor
corresponds to a variable dielectric tensor of vacuum around
gravitational matter if viewed in a flat spacetime. In such a view,
the vacuum possesses a graded refractive index, which slows down the
velocity of light propagating to a gravitational body. This
medium-like property of gravitational space is attributed to the
virtual charge pairs in vacuum. The increasing number density of
virtual dipoles towards to the gravitational centre naturally
suggests a polarizable vacuum interpretation of the gravitational
force. In addition, a dielectric tensor of vacuum could be in a more
general form than that with equal diagonal elements, which makes it
possible that a vacuum under certain conditions turns to be
anisotropic. So, great importance is attached to the observations or
experiments searching for anisotropic propagation of light in a
vacuum modified by specific gravitational or electromagnetic fields.

\ \

\appendix{\textbf{Acknowledgments}}

\

This work was supported by Hangzhou Dianzi University (grant no.
KYS075608069).


\appendix{\textbf{}}


\begin{thebibliography}{00}




\bibitem{Einstein1920}
A. Einstein, \emph{\"{A}ther und Relativit\"{a}tstheorie}
(Springer-Verlag, Berlin, 1920).

\bibitem{Eddington1920}
A. S. Eddington, \emph{Space, Time and Gravitation} (Cambridge
University Press, Cambridge, 1920).

\bibitem{Wilson1921}
H. A. Wilson, Phys.\ Rev. \textbf{17}, 54 (1921).

\bibitem{Dicke1957}
R. H. Dicke, Rev.\ Mod.\ Phys. \textbf{29}, 363 (1957).

\bibitem{Milonni1994}
P. W. Milonni, \emph{The Quantum Vacuum: An Introduction to Quantum
Electrodynamics} (Academic Press, New York, 1994).

\bibitem{Peskin2006}
M. E. Peskin and D. V. Schroeder, \emph{An Introduction to Quantum
Field Theory} (World Publishing Corp., Beijing, 2006).

\bibitem{Puthoff2002}
H. E. Puthoff, Found.\ Phys. \textbf{32}, 927 (2002).

\bibitem{Vlokh2004}
R. Vlokh, Ukr.\ J.\ Phys.\ Opt. \textbf{5}, 27 (2004).

\bibitem{Vlokh2005}
R. Vlokh and M. Kostyrko, Ukr.\ J.\ Phys.\ Opt. \textbf{6}, 120;
\textbf{6}, 125 (2005).

\bibitem{Vlokh2007}
R. Vlokh and O. Kvasnyuk, Ukr.\ J.\ Phys.\ Opt. \textbf{8}, 125
(2007).

\bibitem{Puthoff2005}
H. E. Puthoff, E. W. Davis, and C. Maccone, Gen.\ Rel.\ Grav.
\textbf{37}, 483 (2005).

\bibitem{Vlokh2006}
R. Vlokh and M. Kostyrko, Ukr.\ J.\ Phys.\ Opt. \textbf{7}, 179
(2006).

\bibitem{Ye2008-01}
X. H. Ye and Q. Lin, J.\ Mod.\ Opt. \textbf{55}, 1119 (2008).

\bibitem{Ye2008-02}
X. H. Ye and Q. Lin, J.\ Opt.\ A: Pure Appl.\ Opt. \textbf{10},
075001 (2008).

\bibitem{Weinberg1972}
S. Weinberg, \emph{Gravitation and Cosmology} (John Wiley and Sons,
New York, 1972).

\bibitem{Ye2009-01}
X. H. Ye, arXiv: 0902.1305v1 (2009).

\bibitem{Preuss2005}
O. Preuss, S. K. Solanki, M. P. Haugan, and S. Jordan, Phys.\ Rev.\
D. \textbf{72}, 042001 (2005).

\bibitem{Zavattini2006}
E. Zavattini \emph{et al}., Phys.\ Rev.\ Lett. \textbf{96}, 110406
(2006).









\end{thebibliography}
\end{document}